\documentstyle[twoside,fleqn,psfig,hyperon99_paper]{article}

\newcommand{\AmS}{{\protect\the\textfont2
  A\kern-.1667em\lower.5ex\hbox{M}\kern-.125emS}}

\title{Is the riddle of the hyperon polarizations solved?}

\author{J. Soffer \address{Centre de Physique Th\'eorique, CNRS,\\
        Luminy Case 907, F13288 Marseille Cedex 09, France}}%
        
\begin{document}

\begin{abstract}
We review in this talk some aspects of the exciting field of hyperon
polarization phenomena in high energy reactions, over the last twenty years
or so. On the experimental side, a large amount of significant polarization
data for hyperon and antihyperon inclusive production, has been accumulated
in a rather broad energy range. Many theoretical attempts to explain that
have been proposed and we will discuss some of them, showing their strong 
limitations in most cases. 
\end{abstract}

\maketitle

\section{INTRODUCTION}
Naively it seems reasonable to expect no polarization effect in a 
one-particle inclusive reaction
\begin{equation}
a+b \rightarrow c +X~,
\end{equation}
since one is summing over many different inelastic channels $X$, which should have polarizations
of random magnitudes and signs, such that the sum will average to zero. 
In reality the experimental situation is not so simple and single spin asymmetries have been observed
in many specific reactions and we will first give a rapid tour of these data, mainly for hyperon production. Next we will present some 
current theoretical ideas, which have been proposed to explain the main features of the data. Finally, we will say a few words on 
future expectations and will make our closing remarks.

\section{A RAPID TOUR OF THE HYPERON POLARIZATIONS DATA}
There is a large amount of very significant data for hyperon and antihyperon inclusive production \cite{Pon85,Hel97,Pon99,Smi99}, some
 of them exhibiting a simple pattern, which may help to uncover the underlying particle production mechanism. This 
 polarization effect was first discovered in 1976 at FNAL by studying hyperons produced by a $300~GeV/c$ proton beam on a 
 Beryllium target \cite{Bun76} and it was found that the $\Lambda 's$ produced in the beam fragmentation region have a large polarization 
 perpendicular to the production plane. Since then, many different experiments have collected high statistics data on $\Lambda$  
 inclusive production, which makes it the best known hyperon inclusive reaction.
 
 Let us briefly recall the main characteristics of these proton induced data, which exhibit some interesting regularities \cite{Pon85}:\\
 
 i) the invariant cross section $Ed^3\sigma/dp^3$ depends, to a good approximation, only on $x_F^{\Lambda}$, the fraction of 
 incident proton momentum carried by the $\Lambda$ in the beam direction (in the center of mass ({\it c.m.}) system), and 
 $p^{\Lambda}_T$, the $\Lambda$ transverse momentum, and does {\it not} depend on the {\it c.m.} energy $\sqrt s$.\\
 ii) the transverse polarization $P_{\Lambda}$ is {\it negative} with respect to the direction $\vec n = \vec p_{\it inc} \times \vec p_{\Lambda}$.\\
 iii) $P_{\Lambda}$ is almost energy and target independent for an incident energy ranging from $12~GeV/c$ on a Tungsen target 
 \cite{Abe83} up to $2000~GeV/c$ at ISR \cite{Smi87}.\\
 iv) for $p_T^{\Lambda}$ below $1~GeV/c$ or so, the magnitude of $P_{\Lambda}$ is approximately linear in $p_T^{\Lambda}$, with a 
 slope increasing with $x_F^{\Lambda}$.\\
 v) for $p_T^{\Lambda}$ above $1~GeV/c$, the magnitude of $P_{\Lambda}$ is independent of $p_T^{\Lambda}$, up to $p_T^{\Lambda} \sim 3.5~GeV/c$ and
 approximately linear with $x_F^{\Lambda}$ \cite{Lun89}.
 
 We also have data on other hyperon polarizations at FNAL energy, where one observes, with respect to the $\Lambda$ polarization, 
 an effect of opposite sign for $\Sigma^{\pm}$ and same sign for $\Xi^-$ and $\Xi^0$ \cite{Wil87}. However it seems that $P_{\Xi^-}$ does 
 {\it not} increase with energy, whereas $P_{\Sigma ^+}$ decreases with energy \cite{Hel97}.
 
 Finally, the situation of the antihyperon polarizations is very puzzling since, on the one hand, $P_{\Lambda} \sim P_{\Xi^0} \neq 0$ and
 $P_{\bar \Lambda} \sim P_{\bar \Xi^0} =0$, but on the other hand, $P_{\bar \Sigma^{-}} \sim P_{\Sigma^+}$ and 
 $P_{\bar \Xi^{+}} \sim P_{\Xi^{-}}$ \cite{Hel97},\cite{Erw99}.
 Needless to say that all these peculiarities of the data constitute a real challenge for the theory, 
 some aspects of which we start discussing now.

\section{SOME THEORETICAL IDEAS FOR SINGLE TRANSVERSE SPIN ASYMMETRIES}
Let us consider the reaction (1), where one observes the transverse polarization
state of one of the hadrons (initial or final). The simplest measurable quantity
is the single transverse spin asymmetry (or up-down asymmetry), defined as, for example 
if $c$ is polarized
\begin{equation}
P_{c}=\frac{d\sigma_{c}^{\uparrow} - d\sigma_{c}^{\downarrow}}{d\sigma_{c}^{\uparrow} + d\sigma_{c}^{\downarrow}}~,
\end{equation}
also usually called the $c$ polarization. The transverse spin asymmetry related to the initial particle ( $a$ or $b$)
is called the analyzing power and is denoted by $A_N$. By using the generalized optical theorem, one can write
\begin{equation}
P_{c}d\sigma=Im[f^*_{+}f_{-}]~,
\end{equation}
where $d\sigma = d\sigma_{c}^{\uparrow} + d\sigma_{c}^{\downarrow}$ is the corresponding unpolarized
inclusive cross section. It is described by means of $f_{+}$, the forward {{\it non-flip} $3\rightarrow3$
helicity amplitude $ab\bar{c}_{\lambda} \rightarrow ab\bar{c}_{\lambda}$, where $\lambda=\pm $ is the same
on both sides. Moreover $f_-$ is the forward {\it flip} amplitude $ab\bar{c}_{\lambda} \rightarrow ab\bar{c}_{-\lambda}$. 
In order to get a non-vanishing $P_c$ (or $A_N$), one needs, a non-zero $f_-$ and furthermore it should have
a phase difference with $f_+$. This point is important and should be taken seriously, if we want to have
a real understanding of the available experimental data. It is another way to say that a non-zero $P_c$
corresponds to a non-trivial situation, which reflects a high coherence effect among many different inelastic
channels. In principle, in addition to the {\it c.m.} energy $\sqrt s$, all these observables are expected 
to depend on two kinematic variables defined as  $x_F^c=2p_L^c/\sqrt s$ and $x_T^c=2p_T^c/\sqrt s$, where $p^c_L$
and $p^c_T$ are the {\it c.m.} longitudinal and transverse momentum of $c$ with respect to the incident beam direction. 
Clearly one has the kinematic limits $-1\leq x^c_F \leq +1$ and $0 \leq x^c_T \leq+1$ and one should distinguish
two different kinematic regions:\\

i) The beam fragmentation region\\
It corresponds to a region where $c$ carries a sizeable $x^c_F$ say, $0.3\leq x^c_F \leq 0.8$, with a small value
of $x^c_T$ say, $0 \leq x^c_T \leq 0.1-0.15$. Similarly, one can consider the target fragmentation region, corresponding
to $x^c_F$ of opposite sign.\\

ii) The hard scattering region\\
It corresponds to a region where $c$ carries a sizeable $x^c_T$ say, $x^c_T \geq 0.15$, with $x^c_F \sim 0$.\\

Rather different dynamical mechanisms are expected to be at work in these different kinematic regions and since most
of the hyperon polarization data are in the fragmentation region, we will first consider it.

\subsection{The fragmentation region}
Actually, in this region the polarization mechanism is essentially based on a soft process, where perturbative
QCD does not apply.
There are {\it two} classes of dynamical models available in the literature which will be now discussed below.\\

{\it a) Semiclassical models}\\
These models provide simple arguments for a qualitative description of the hyperon polarizations, but since they
fully ignore the relevance of the phase difference, which is crucial, as mentioned above, they are unable to make
solid quantitative predictions.\\ 

 a1) - The Lund model

We recall that in terms of the constituent quarks, the proton beam fragmentation into a $\Lambda$ with a $p_T^{\Lambda} 
\neq 0$, corresponds to the replacement of a valence $u$ quark, in the projectile, by a strange quark $s$ coming from
the sea, which must be accelerated along the beam direction, and acquiring also a non-zero $p_T$. Moreover we will be assuming
a $SU(6)$ wave function, where the $(ud)$ system of the $\Lambda$ is in a singlet state, so the $\Lambda$ polarization
is that of the $s$ quark. In the Lund model \cite{And83}, an incoming $(ud)$ diquark with spin $S=0$ and isospin $I=0$,
stretches the confined color field in the collision region and a $s\bar s$-pair is produced. The $s$ quark is needed
to make the final $\Lambda$ and since $p_T^{\Lambda} \neq 0$, one assumes that part of this transverse momentum is
provided by the $s$ quark, which has to be compensated by that of the $\bar s$ quark. As a result, the $s\bar s$-pair
has an orbital angular momentum which is assumed to be balanced by the spin of the $s\bar s$-pair. From this mechanism, one 
expects in proton induced reactions, a {\it negative} $\Lambda$ polarization increasing with $p_T^{\Lambda}$, in agreement
with the sign of the data but whose magnitude is difficult to predict. The $\Sigma$ polarization follows from the knowledge 
of the $SU(6)$ wave functions and since for the $\Sigma$, the $(ud)$ diquark has a spin $S=1$, it is natural to expect
an opposite polarization, in accordance with the data. However this simple picture cannot be correct and 
we are now indicating, several drawbacks of the model:\\
- The case of the $\Xi$ requires the production of two strange quarks and involves additional assumptions in the model,
 which does not explain why $\Lambda$ and $\Xi$ have nearly equal polarizations. In addition, nothing is said about the
description of the unpolarized cross sections, observables of crucial importance to pin down the dynamics and which are very different 
for $\Lambda$ and $\Xi$ production.\\
- The model does not suggest any $x_F$ dependence of the polarization.\\
- In this approach, $P_{\Lambda}$ increases linearly with $p_T^{\Lambda}$, but one does not know why it stops growing,  
which is needed since $|P_{\Lambda}| \leq 1$ and moreover, as recalled above, because the data saturates for 
$p_T^{\Lambda}$ above $1 GeV/c$.\\
- As we have seen, the $s$ quark and the $\bar s$ quark move into opposite directions. So for $\Lambda$
production, by looking at the final state $K^+\Lambda$, since $\vec K^+$, the direction of the $K^+$, gives the direction of the $\bar s$,
one should observe that $P_{\Lambda} \neq 0$, only when $K^+$ and $\Lambda$ are in opposite hemispheres. The E766 experiment
at BNL, with a proton beam of $27.5~GeV/c$ \cite{Fel96}, has made extensive studies of an {\it exclusive} channel
$pp\rightarrow p \Lambda K^+ \pi^+ \pi^-$ and also other channels with more than one $(\pi^+ \pi^-)$ pair produced \cite{Fel99}. They found that
$P_{\Lambda}$ can be parametrized, in a limited kinematic region, according to
\begin{equation}
P_{\Lambda}(x_F^{\Lambda},p_T^{\Lambda})=(-0.443 \pm 0.037)x_F^{\Lambda}\cdot p_T^{\Lambda}~,
\end{equation}
in agreement with the inclusive data. However they found no correlation between $\vec K^{+}$ and the value of
$P_{\Lambda}$.

At this point, it is worth to emphasize the importance of exclusive channels, in particular the simple diffractive reaction
$pp\rightarrow p (\Lambda K^+)$, for which a large negative $P_{\Lambda}$  $(\sim -60\%)$ has been observed at the
ISR \cite{Hen92}, consistent with a very recent result from the experiment E690 \cite{Chr99}. There is no
indication for such a correlation between $\vec K^+$ and $P_{\Lambda}$ and it is conceivable to try
to relate this large value of $P_{\Lambda}$  to a diffractive mechanism with Pomeron exchange, since one observes no
energy dependence between $\sqrt s = 40~GeV$ and  $62~GeV$.\\

a2) - The recombination model

This is another approach \cite{Deg86} based on semiclassical arguments, which are applied to a recombination mechanism. As we have
already pointed out, to make a $\Lambda$ from the fragmentation of an incident proton, one needs to recombine a fast $(ud)$ diquark
from the proton with a slow $s$ quark from the sea. If $\vec F$ denotes the unspecified color force which gives this acceleration, 
the $s$ quark of velocity $\vec v$ feels the effect of the Thomas precession given by $\vec \omega_T \sim \vec F \times \vec v$, which
has the direction of the normal to the hadronic scattering plane $\vec n = \vec p_{\it inc} \times \vec p_{\Lambda}$. In order to minimize
the energy $\vec S \cdot \vec \omega_T$ associated to this effect, the spin $\vec S$ of the $s$ quark must be opposite to 
$\vec \omega_T$, so this leads to expect a negative $\Lambda$ polarization in $pp \rightarrow \Lambda X$. Although this approach
is different from the Lund model, it seems to lead to similar observable effects. However, one direct consequence of the recombination
model is that, in the inclusive reaction $K^-p\rightarrow \Lambda X$, if the $\Lambda$ is produced in the $K^-$ beam fragmentation,
the $s$ quark needed to make the $\Lambda$, is now coming from the $K^-$. It is fast and has to be decelerated when joining a $(ud)$ diquark
 from the sea, so the sign of $P_{\Lambda}$ is reversed. This is in agreement with the data \cite{Gou86}, but the magnitude
 of $P_{\Lambda}$ in a $K^-p$ collision is approximately twice as large as in a $pp$ collision, so clearly the simplest version
 of the model does not explain this big difference. Let us recall that this large $P_{\Lambda}$ in $K^-p$ collision is also energy
 independent between $p_{\it inc} =12~GeV/c$ \cite{Arm85} and $176~GeV/c$ \cite{Gou86}, an interesting scaling property.
  Another strong prediction of this model
 is that the polarization of the antihyperons must be zero. This is in agreement with the data for $\bar \Lambda$, but at variance
 with some other cases, as we recalled above. Moreover the Thomas precession does not act exclusively on the strange quark, but also
 on the diquark and depends on its possible spin states $j=0$ and $j=1$. Consequently, one has to introduce more parameters
 to make relative predictions for a fair number of inclusive reactions, a situation not very satisfactory.\\
 In addition to the polarization $P$, we have seen that one can consider the analyzing power $A_N$ and also
 another spin-observable, the spin transfer
 parameter denoted by $D_{NN}$, which measures the fraction of the transverse component of the beam polarization transferred to the hyperon. Another
 bad failure of the model is that, it predicts $A_N = D_{NN} =0$, in strong contradiction with the results of the E704 experiment at
 FNAL. They found in $pp$ inclusive $\Lambda$ production at $200~GeV/c$, a substantial negative $A_N$, at relatively large $x_F^{\Lambda}
 (\geq 0.5)$ \cite{Bra95}, and a positive $D_{NN}$ up to about $30\%$, also at high $x_F^{\Lambda}$ \cite{Bra97}.\\
 
 a3) - The Berlin model
 
 In this model \cite{Bor96,Lia97}, one is claiming that the existence of striking analyzing power $A_N$ observed in $\pi^{\pm}$
 inclusive production at FNAL \cite{Ada91}, is a strong indication for orbiting valence quarks in a polarized proton. So the orbital
 motion should be taken into account with the following semiclassical picture:\\
 A hadron is polarized, {\it if and only if}, its valence quarks are polarized and due to a significant {\it surface effect}, only
 valence quarks retain the information about polarization.
 
 This leads to conclude that, for example, a meson which is formed by {\it direct fusion} of an upward valence quark with an antiquark from the sea, gets 
 a transverse momentum from the orbital motion of the valence quark, to the left, looking downstream. So in $pp \rightarrow \pi^+ X$, since
 $\pi^+= (u_v\bar d_s)$ one should have $A_N>0$, whereas in $pp \rightarrow \pi^- X$, one should have $A_N<0$, which are both in agreement with
 the data \cite{Ada91}. In order to make more quantitative statements, they must assume that the $x_F$ of the produced hadron ($\pi^+$ or $\pi^-$)
 is that of the initial polarized valence quark $x$, and use the information one has on the polarized quark distributions $\Delta u_v(x)$ and 
 $\Delta d_v(x)$, obtained from polarized deep inelastic scattering. Of course the model makes no statement about the $p_T$ dependence.
 From the quark structure of the $K$ mesons, one predicts the same $A_N$  for $K^+$ and $\pi^+$ inclusive production and also the same 
 for $K^0_s$ and $\pi^-$. The first prediction remains to be checked but there is a good indication that the second one is correct from
 the AGS data \cite{Bon90}. Another consequence is that $A_N$ is zero for $K^+$ and $\bar K^0_s$ inclusive production and this last prediction also, 
 has not yet been verified.\\
 Let us now return to hyperon inclusive production and more specifically to the $\Lambda$ case. There are three 
 possibilities for the direct formation: the incident proton can release either a $(u_vd_v)$ diquark, or a $u_v$ quark or a $d_v$ quark,
 which must be combined with the appropriate missing piece from the sea, to make a $\Lambda$. The calculations lead to a reasonable agreement with the data
 for $P_{\Lambda}$, $A_N$ as shown in refs.\cite{Bor96,Lia97}, and $D_{NN}$ is predicted to be positive in accordance with \cite{Bra97}.
 To summarize, this model has some predictive power, but contains several key assumptions, which make it not fully convincing. Moreover, nothing is
 said about the other hyperons and neither about the puzzling situation of the antihyperons.\\
 
 At this stage we will make a short digression on some positivity conditions, which are not necessarily well known. One can show
 that for any inclusive reaction of the type eq.(1), where $a$ and $c$ are any polarized spin-1/2 particles and $b$ is an unpolarized particle,
 of any spin, one has \cite{Don72} the following very general constraints among $P_c$, $A_N$ and $D_{NN}$
 \begin{equation}
 1 \pm D_{NN} \geq | P_c \pm A_N|~.
 \end{equation}
 These inequalities, which are model independent rigorous conditions, 
  must be satisfied for any kinematic values of the variables $x^c_F$, $p_T^c$ and $\sqrt s$.\\
 As an example, we have tested the results of the E704 experiment at $p_{\it inc}=200~GeV/c$:
 
 for $p_T^{\Lambda} \sim 1~GeV/c$ 
 and $x_F^{\Lambda} \sim 0.8$, one has $D_{NN} \sim 30\%$, $A_N \sim -10\%$ and $P_{\Lambda} \sim -30\%$, showing that the above
 constraints are indeed well satisfied.\\

 {\it b) Regge type models}\\
 It is important to try to generate or to justify the origin of the phase difference between ${\it f_+}$ and ${\it f_-}$ occuring in eq.(3)
 and this is what one does, in the two phenomenological models, we are presenting now 
 \footnote{ There is also an attempt to produce the phase using one-loop diagrams in perturbative QCD and 
 the recombination of polarized quarks to form polarized hadrons. This hybrid model is described in ref. \cite{Gol99}.}.\\
 
 b1) - The Milano model
 
 In the fragmentation region, this phase might be resulting from final state interactions and more precisely one can invoke
 a dynamical model \cite{Bar92} based on the production of various baryon resonances, in general out of phase, which then decay
 to give the observed $\Lambda$. Unlike the theoretical frameworks discussed so far, here one is trying, first, to provide a good representation
 of the inclusive unpolarized cross section. It is known that hadron fragmentation is well described by the triple-Regge model and
 for example for $\Lambda$ production, they consider three production mechanisms: (a) direct production, (b) intermediate baryon
 dissociation ($\Sigma, \Sigma^*$), (c) $\Sigma^0$ electromagnetic decay. The calculation, which involves the relevant Regge residues,
 leads to a reasonable unpolarized $\Lambda$ spectrum. We note that the direct production contribution produces only unpolarized
 $\Lambda$ and dominates at large $p_T^{\Lambda}$. As a result, the predicted $P_{\Lambda}$, which has the right negative sign and the correct 
 magnitude at low $p_T^{\Lambda}$, tends to decrease at high $p_T^{\Lambda}$, in contradiction with experiment. This has been extended to
 the $\Sigma$ case and although they get a positive sign, they fail to predict the right magnitude of $P_{\Sigma}$. It seems hard to
 use this approach for $\Xi$ production, since it would requiere a rather elaborate extension of the Regge model.\\
 
 b2) - The one-pion exchange model
 
 Here one assumes that $\Lambda$ production in the fragmentation region is dominated by a reggeized one-pion exchange, a model proposed
 several years ago and which gives a successful description of various exclusive and inclusive reactions. As is well known, if
 quantum numbers allow, pion exchange generally dominates hadronic amplitudes, especially at small momentum transfers. Therefore
 the leading contribution involves the diagram such that, the multiperipheral chain reduces only to the binary reaction 
 $\pi p \rightarrow K \Lambda$ and the total $\pi p$ cross section, connected by the exchange of an off-shell reggeized pion \cite{Sof92}.
 The $\Lambda$ spectrum is obtained from the $pp\rightarrow K \Lambda X$ cross section, after integration over the kaon phase space. Note that
 the binary reaction has a subenergy in the resonance region up to $10~GeV/c$ or so. One predicts all the basic features of the
 $\Lambda$ spectrum, including the scaling property, with no free parameter. A crucial test of the model is the calculation of $P_{\Lambda}$,
 which is obviously directly related to the $\Lambda$ polarization of the binary reaction $\pi p \rightarrow K \Lambda$ at fairly low energy.
 It is essentially negative and therefore leads to a negative $P_{\Lambda}$, with a magnitude consistent with the data. 
 It would be also interesting to know what this model predicts for $A_N$ and $D_{NN}$. Although the
 calculation was not done for $P_{\Sigma}$, one can anticipate a positive sign, due to the fact that in the binary reaction
 $\pi^+ p \rightarrow K^+ \Sigma^+$, the polarization is positive. However it seems not possible to extend this approach to
 the $K$ induced reaction $K^- p\rightarrow \Lambda X$, which is unfortunate.
 
 \subsection{The hard scattering region}
 This kinematic region involves short distance interactions, where perturbative QCD is expected to apply.
 Naively, a transverse spin asymmetry in a parton subprocess is anticipated
 to be of the form
 \begin{equation}
 \hat A_N \sim \alpha_s \frac{m_q}{p_T}~,
 \end{equation}
 because the spin-flip amplitude is proportional to the quark mass and the imaginary part (see eq.(3)) must be produced by a
 one-loop diagram, which generates the strong coupling constant $\alpha_s$. Clearly this result, which is valid
 only at the twist-2 level in QCD, is extremely small and will lead also to a small hadronic asymmetry.
 On the experimental side, this kinematic region is hardly accessible, for statistical reasons. On the one hand, there is 
 an indication for a large effect in $\pi^0$ production in $\pi$ induced reactions at $40~GeV/c$ \cite{Apo90} and, on the other hand, 
 from E704 in $pp\rightarrow \pi^0 X$  at $200~GeV/c$ \cite{AAA96}, $A_N$ is consistent with zero. However more then ten years ago,
 a self consistent approach to single spin asymmetries at the twist-3 level in QCD was developed \cite{Efe85}. In order to avoid the
 extra complication coming from the distributions and fragmentation functions, let us consider the simplest, perhaps, academic reaction 
 $\gamma p^{\uparrow} \rightarrow \gamma X$. According to this theoretical approach the spin transverse proton asymmetry has the form
 \begin{equation}
 A_N \sim \frac {M_p b(x_1,x_2)}{p_T}~, 
 \end{equation}
 where the quark mass $m_q$ has been replaced by the proton mass $M_p$ and the other killing factor $\alpha_s$ is now replaced by
 the quark-gluon correlator $b(x_1,x_2)$, a new two-arguments ($x_1$ and $x_2$) structure function, which must be extracted from the data, just
 like any ordinary parton distribution. Of course the remaining $p_T$ in the 
 denominator reflects the fact that we are dealing with a twist-3 effect, which is expected to decrease for very large $p_T$ values. 
 In the cases of $pp \rightarrow \Lambda^{\uparrow} X$ or $pp^{\uparrow} \rightarrow \pi X$,
 discussed before in the fragmentation region, the above result has to be convoluted by the quark distributions and the final hadron
 fragmentation functions, but the key question which remains to be answered is: how large are these quark-gluon correlators? It is 
 an interesting experimental problem which, hopefully, will be solved in the near future, in particular, with the polarized $pp$
 collider at RHIC-BNL, due to start operating very soon. This new unique facility will also allow to get some relevant information
 on the $\Lambda$ polarized fragmentation functions \cite{DeF98}.
 
 \section{CONCLUDING REMARKS}
 We have shown that in hadronic spin physics at high energy, the field of transverse spin asymmetries is extremely rich, in particular,
 by looking of the available data. We are facing a considerable number of polarization effects in hyperon and antihyperon inclusive
 (and exclusive) production, which remain widely unexplained. Therefore we believe it is fair to conclude that, despite several theoretical
 efforts over the last twenty years or so, theory is left behind and has to make urgent progress, to catch up with the puzzling
 experimental situation. If there exists a universal physical picture to shade light on this important area
 of high energy physics, it has not been discovered yet.

{\it Acknowledgements}

It is my pleasure to thank E. Monnier and all the organizers for their invitation and for setting up this excellent workshop 
in such a pleasant and stimulating atmosphere.\\

\end{document}